\documentclass[12pt,preprint]{aastex}
\usepackage{graphicx}

\slugcomment{To appear in the Astrophysical Journal}
\shorttitle{Oxygen X-ray Fluorescence}
\shortauthors{J.J.~Drake \& B.~Ercolano}
\begin{document}

\title{On the Detectability of Oxygen X-ray Fluorescence and its Use 
as a Solar Photospheric Abundance Diagnostic} 

\author{Jeremy J.~Drake\altaffilmark{1}}
\author{Barbara Ercolano\altaffilmark{1}}
\affil{$^1$Smithsonian Astrophysical Observatory,
MS-3, \\ 60 Garden Street, \\ Cambridge, MA 02138}
\email{jdrake@cfa.harvard.edu}

\begin{abstract}
Monte Carlo calculations of the O~K$\alpha$ line fluoresced by coronal
X-rays and emitted just above the temperature minimum region of the
solar atmosphere have been employed to investigate the use of this
feature as an abundance diagnostic.  While quite weak, we estimate
line equivalent widths in the range 0.02-0.2~\AA, depending on the
X-ray plasma temperature.  The line remains essentially uncontaminated
by blends for coronal temperatures $T \leq 3 \times 10^6$~K and should
be quite observable, with a flux $\ga 2$~ph~s$^{-1}$arcmin$^{-2}$.
Model calculations for solar chemical mixtures with an O abundance
adjusted up and down by a factor of 2 indicate 35-60\%\ changes in
O~K$\alpha$ line equivalent width, providing a potentially useful O
abundance diagnostic.  Sensitivity of equivalent width to differences
between recently recommended chemical compositions with ``high'' and
``low'' complements of the CNO trio important for interpreting
helioseismological observations is less accute, amounting to
20-26\%\ at coronal temperatures $T \leq 2\times 10^6$~K.  While still
feasible for discriminating between these two mixtures, uncertainties
in measured line equivalent widths and in the models used for
interpretation would need to be significantly less than 20\%.
Provided a sensitive X-ray spectrometer with resolving power $\geq 1000$ and
suitably well-behaved instrumental profile can be built, X-ray
fluorescence presents a viable means for resolving the solar ``oxygen
crisis''.
\end{abstract}


\section{Introduction}
\label{s:intro}

At the turn of the millennium, precise agreement between the observed
oscillation spectrum of the Sun and predictions of theoretical solar
models, built using the best abundance assessments of the day
\citep{Grevesse.Sauval:98}, represented a triumph of modern
astrophysics \citep[e.g.\ ][]{Bahcall.etal:05}.  This illustrious
accord has recently been clouded by sophisticated re-assessments of
the solar complement of light elements based on 3-D non-LTE
hydrodynamic photospheric modelling: such models demand less of the
elements C, N, O and Ne that are important for the opacity of the
solar interior by 25-35~\%\ compared to earlier assessments
\citep{Asplund.etal:05}.  Solar models employing this chemical
composition lead to predictions of the depth of the convection zone,
helium abundance, density and sound speed in serious disagreement with
helioseismology measurements
\citep{Basu.Antia:04,Turck-Chieze.etal:04,Bahcall.etal:05}.  This
state of affairs is now commonly referred to as the ``solar model
problem'', or the ``solar oxygen crisis''
\citep[e.g.\ ][]{Ayres.etal:06}.

To solve this problem, \citet{Antia.Basu:05} and \citet{Bahcall.etal:05c}
suggested the uncertain solar Ne abundance might be raised to
compensate for lower C, N and O abundances.  \citet{Drake.Testa:05}
found empirical support for this in {\it Chandra} high resolution
X-ray spectra of mostly magnetically active stars for which the Ne/O
abundance ratio appears consistently higher by a factor of $\sim 2$ or
more than the currently recommended solar value of Ne/O=0.15 by
number.  They argued that existing solar coronal Ne/O measurements
supporting the latter \citep[see,
e.g.,][]{Drake.Testa:05,Schmelz.etal:05,Landi.etal:07} might be biased
by chemical fractionation.  However, subsequent studies of the
helioseismology data using different theoretical models disagree as to the
viability of such a solution \citep{Delahaye.Pinsonneault:06,
Zaatri.etal:07,Lin.etal:07}.

An alternative possibility is that the suggested downward C, N and O
abundance revisions are unwarranted.  \citet{Ayres.etal:06} favour the
\citet{Grevesse.Sauval:98} value of O/H$=8.83$ based on
multi-component atmospheric modelling and analysis of CO lines.
However, a very recent 3D empirical study of spatially-resolved
spectropolarimetric observation of Fe~I and O~I lines by
\citet{Socas-Navarro.Norton:07} finds O/H$=8.63\pm 0.1$, in good
agreement with \citet{Asplund.etal:05}.

The difficulty in determining the solar O abundance lies in the
complexity of the non-local thermodynamic equilibrium optically-thick
line formation problem in the inhomogeneous and turbulent photosphere.
In this paper, we study a possible alternative approach.
\citet{Drake.Ercolano:07} recently investigated the utility of the
weak photospheric K$\alpha$ fluorescence line of neutral Ne excited by
coronal X-ray photoionisation as an independent abundance diagnostic.
Here, we turn our attention to the O~K$\alpha$ line at
23.37~\AA\ (530.6~keV). In contrast to the complex photospheric
absorption line formation problem, the formation of the X-ray
fluorescent O~K$\alpha$ line is relatively simple, depending only on
the geometry of the overlying coronal source of irradiation and the
heliocentric angle of the photospheric region in question
\citep[e.g.][]{Bai:79,Drake.etal:07,Drake.Ercolano:07}.  We present
Monte Carlo calculations of the fluorescent O~K$\alpha$ line in \S2,
and address its potential as an abundance diagnostic in \S3 and \S4.

\section{Monte Carlo Calculations of Oxygen Fluorescence}
\label{s:monte}

We adopt the same computational methods described in earlier work on
Fe and Ne fluorescence lines, and the reader is referred to
\citet[hereafter Paper~I]{Drake.etal:07} and \citet[Paper
  II]{Drake.Ercolano:07} for details.  A brief summary is provided
below.

X-ray ``characteristic'' (fluorescence) K$\alpha$ lines correspond to
the $2p-1s$ decay of the excited state resulting from ejection of an
inner-shell $1s$ electron in the neutral (or near-neutral) atom by either
electron impact or photoionisation.  In the case of solar and stellar
photospheric fluorescent lines, the former is expected to dominate
\citep{Basko:78,Bai:79,Parmar.etal:84}.  For a given coronal X-ray
source spectrum, $F(\lambda)$, the observed flux of K$\alpha$ photons
from the photosphere then depends on the photospheric abundance $A$ of the
fluorescing species relative to that of other elements of significance
for the photoabsorption opacity in the vicinity of the $1s$ ionisation
edge; the height $h$ of the emitting source; and the heliocentric
angle $\theta$ between the emitting source and the observer
\citep{Bai:79}.

Fluorescent lines are formed in the region of the atmosphere
corresponding to optical depth of approximately unity for the primary
K-shell ionising photons.  In the case of O, for normal incidence the
solar atmospheric Model C (VALC) of \citet*{Vernazza.etal:81} reaches
the K-shell photoabsorption $\tau=1$ depth slightly above the
temperature minimum, at a gas temperature of about 5600~K and about
800~km above the point where the continuum optical depth at 5000~\AA,
$\tau_{5000}$, is unity.

The expected intensity of the emergent O~K$\alpha$ line was computed
using a modified version of the 3D Monte Carlo radiative transfer code
MOCASSIN \citep[][see also Paper I for a specific description of
  fluorescence calculations]{Ercolano.etal:03,Ercolano.etal:05}.  This
code follows energy packets representing incident X-ray photons as
they interact with the photospheric gas through photoelectric
absorption or scattering events until the packets escape.
O~K$\alpha$ packets are produced following K-shell X-ray absorption
with a probability dictated by the K$\alpha$ fluorescent yield for
oxygen: these packets are followed in the same fashion.

We adopted a value of $8.3\times 10^{-3}$ for the fluorescence yield
of neutral oxygen \citep{Krause:79}.  Coronal spectra used for the
incident X-ray radiation field were computed using emissivities from
the CHIANTI database \citep{Landi.etal:06} and the ion populations of
\citet{Mazzotta.etal:98}, as implemented in the PINTofALE\footnote{The
  Package for INTeractive Analysis of Line Emission is freely
  available from http://hea-www.harvard.edu/PINTofALE/} IDL suite of
programs \citep{Kashyap.Drake:00}.  

For a given coronal X-ray spectrum, the maximum intensity of a
fluorescent line is achieved for a heliocentric angle $\theta=0$ and
height $h=0$.  As in the case of the earlier study of Ne, since we are
primarily interested here in the observability of the fluorescent
line, we adopt $\theta=0$ and $h=0$ as baseline parameters, these
values yielding the maximum possible line strength.

The wavelength of the O~K$\alpha$ doublet has not been determined with
high precision and we adopted the value $23.37\pm 0.02$~\AA,
corresponding to the mean of values found for the energy level of the
excited $[1s]2p^4\,^4P$ of $530.6\pm 0.3$~eV from laboratory
measurements by \citet{Krause:94}, \citet{Caldwell.etal:94} and
\citet[][see also the review by
  \citealt{Garcia.etal:05}]{Stolte.etal:97}.  This wavelength is in
good agreement with the position of the $2p^3\,^4S \rightarrow
[1s]2p^4\,^4P$ resonance seen in absorption in the interstellar medium
toward bright X-ray continuum sources by \citet{Juett.etal:04}.

The O~K$\alpha$ flux was computed for isothermal irradiating coronal
spectra with plasma temperatures in the range $10^6$-$10^7$~K and
coronal and photospheric 
chemical compositions of \citet[][hereafter GS; with O/H=8.83 on the
  usual log+12 
  scale]{Grevesse.Sauval:98} and \citet[][O/H=8.66]{Asplund.etal:05}.
As noted in \S1, these two chemical mixtures differ not only in O
abundance but also, primarily, in C, N and Ne abundances: these
differences turn out to be important for the O fluorescent line and
are discussed in more detail below.  To investigate changes in the O
abundance only, we also performed calculations using a photospheric 
composition corresponding to the 
GS abundance mixture with O elevated and decreased by a
factor of 2 (O/H=9.13,8.53, or [O/H]=$\pm0.3$).


\section{Strength of the O K$\alpha$ Line}
\label{s:okstrength}

\subsection{Monte Carlo Results}
\label{s:mcresults}

\begin{figure}
\begin{center}
\includegraphics[width=1.0\textwidth]{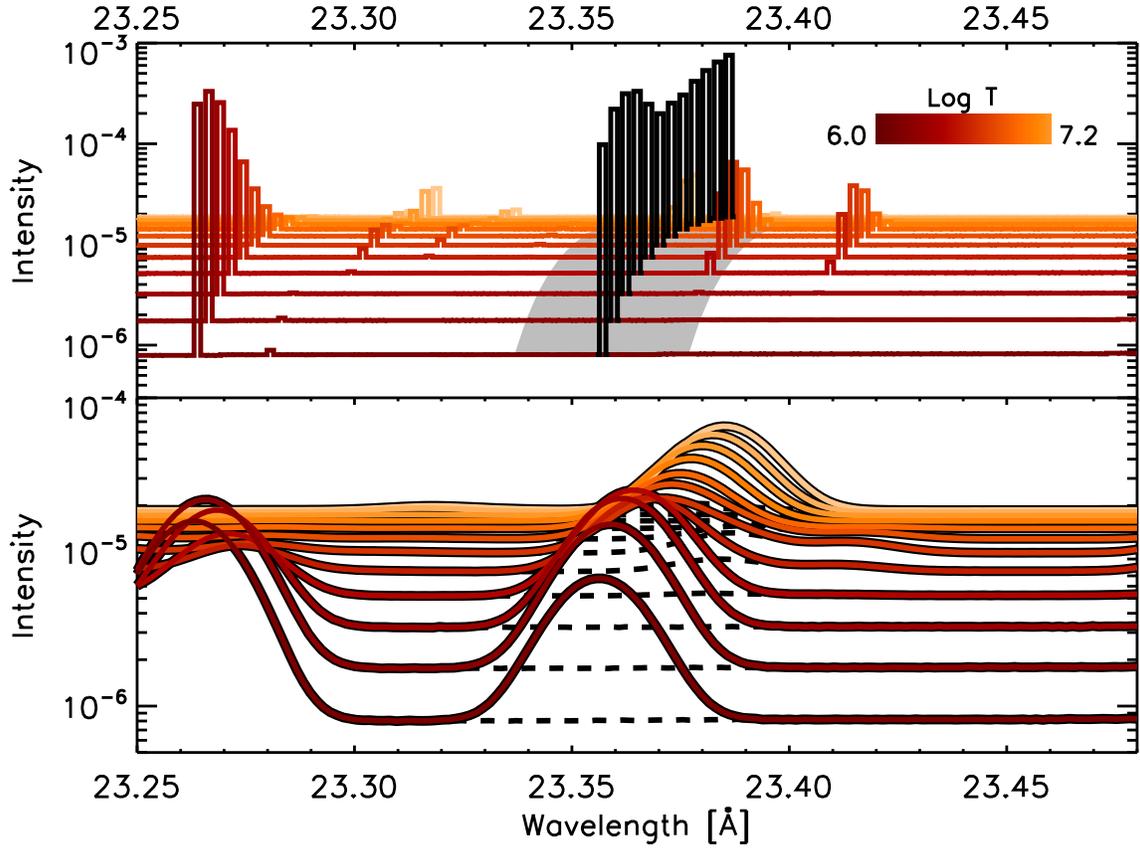}
\end{center}
\caption{Top: The strength of the O~K$\alpha$ shown in comparison to
  neighbouring and blending lines in the fluorescing coronal X-ray
  spectrum for coronal isothermal plasma temperatures in the range
  $10^6$--$10^{7.2}$~K (wavelengths shown off-set for clarity).  The
  underlying grey shaded region represents the uncertainty in the
  wavelength of the O~K$\alpha$ line.  Bottom: The same coronal
  spectra with (solid curves) and without (dashed) the addition of
  O~K$\alpha$ smoothed to a resolving power (FWHM) of
  $\lambda/\Delta\lambda=1000$.  }
\label{f:spectra}
\end{figure}

The photospheric O~K$\alpha$ fluorescent line is shown in comparison
with the direct coronal spectrum for the case of GS abundances for
both coronal spectrum and photosphere  
in Figure~\ref{f:spectra} for the
range of coronal plasma temperatures $6.0 \leq \log T \leq 7.2$.  The
lower panel illustrates the same spectra seen at a resolving power of
$\lambda/\Delta\lambda=1000$, where $\Delta\lambda$ is assumed to be
the full-width at half-maximum of a Gaussian instrument response
function.  Unlike the Ne~K$\alpha$ line, the O~K$\alpha$ transition
forms in a relatively uncrowded spectral region and the CHIANTI
database predicts no devastatingly large blends in the immediate vicinity of
23.37~\AA.  For temperatures approaching $10^7$~K, a line of Fe~XXIII
at 23.36~\AA\ appears, but for GS abundances is always dominated by
the O~K$\alpha$ line by a factor of 10.  Lying redward of our adopted
line centre is Ca~XV~$\lambda 23.39$, which peaks at slightly cooler
temperatures ($\log T\sim 6.6$) at an intensity of about 30~\%\ of
that of O~K$\alpha$.  

The full list of lines in the CHIANTI database within $5\sigma$ of the
O~K$\alpha$ wavelength range ($\pm 0.05$~\AA\ from $23.37\pm
0.02$~\AA) with intensities $\ge 10^{-4}$ times that of the brightest
line (Fe~XXIII~$\lambda 23.363$) are listed in Table~\ref{t:5sigma}.
The NIST Atomic Spectra Database (version 3.1.2;
\citealt{Ralchenko.etal:07}) also lists two other transitions within
the $5\sigma$ range from Ti~XIII and Sc~XIV.  Neither of these are
expected to be of any significance since their solar abundances
according to the GS assessment are 300 and 21000 lower than that of
Fe, respectively.

\begin{figure}
\begin{center}
\includegraphics[width=0.80\textwidth]{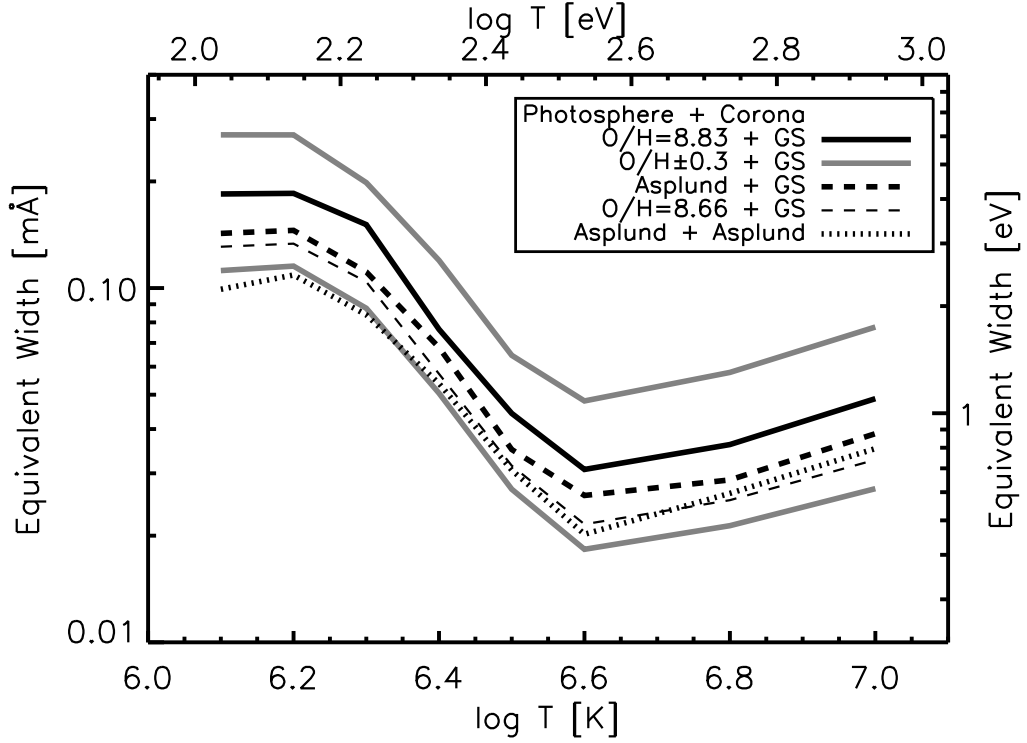}
\end{center}
\caption{The equivalent width of the O~K$\alpha$ line with respect to
  the ionising coronal X-ray spectrum for heliocentric angle
  $\theta=0$ as a function of isothermal plasma temperature, computed
  for different coronal and photospheric chemical compositions: GS
  corona + GS photosphere (solid black curve), GS corona + GS
  photosphere with the O abundance differing by factors of 2 (solid
  grey), GS corona + \citet{Asplund.etal:05} photosphere (heavy
  dashed), GS corona + GS photosphere with \citet{Asplund.etal:05} O
  abundance (O/H=8.66; light dashed); and \citet{Asplund.etal:05}
  corona + \citet{Asplund.etal:05} photosphere (black dotted).  }
\label{f:ew}
\end{figure}

The equivalent width (EW) of the O~K$\alpha$ line relative to coronal
fluorescing spectra with GS abundances is illustrated in
Figure~\ref{f:ew} for a range of isothermal plasma temperatures and
different photospheric O abundances.  The Monte Carlo sampling error
on the computed EWs is estimated to be no larger than 7\%.  One
striking feature of the trend of EW with $T$ for all compositions is
the sharp decline with rising temperature for $\log T\la 6.6$: this
contrasts with both Ne~K$\alpha$ and Fe~K$\alpha$, whose EWs exhibit a
steady, monotonic rise with increasing $T$ (Papers~I \& II),
corresponding to a commensurate increase in the number of ionising
photons.  The different behaviour of O~K$\alpha$ arises because, for
coronal plasmas with $\log T\la 6.6$ and solar composition, K-shell
photoionisation of oxygen is due mostly to line rather than continuum
radiation.  Toward hotter temperatures, the important line-emitting
photoionising species, including O~VII and O~VIII, become ionised, and
continuum contributions begin to dominate.

Regarding sensitivity to O abundance, Figure~\ref{f:ew} illustrates
that the EW changes by a smaller factor than expected based on a
proportional relation with the photospheric the O abundance.  As
discussed by \citet{Bai:79} and in Paper~I, this is a result of the
the O~K-shell photoionisation cross-section being a significant
component of the total opacity near threshold: for very large O
abundances where O begins to dominate the opacity, the EW will tend to
a constant value dictated simply by equipartition of ionising photons
between O K- and L-shells.  Nevertheless, for ``normal'' ranges of O
abundance the O~K$\alpha$ line variations are significant: we find
typical changes of 35-60\%\ over the full range of coronal
temperatures for O abundance variations of a factor of 2.  


When the other abundant light elements C and N are allowed to scale
with the O abundance the diagnostic fares slightly less well but does
provide discrimination between the photospheric compositions of GS and
\citet{Asplund.etal:05}.  The \citet{Asplund.etal:05} O abundance is
48\%\ lower than that of GS, and the O~K$\alpha$ EWs differ by
20-25\%.  For a given exciting X-ray spectrum, changes in fluorescent
line strength with different photospheric parameters depend on changes
in the {\em fraction of ionising photons that ionise the parent shell}
of the fluorescent line in question.  For a solar composition, two
other sources of opacity in the vicinity of the O edge are C and N.
The chemical compositions of GS and \citet{Asplund.etal:05} differ by
$< 10$\%\ in O/C and O/N ratios, and the lockstep changes in these
elements that dilutes the effect of the different O abundances
relative to H on the O~K$\alpha$ EW.  Despite this slightly lower
sensitivity of the fluorescent line to the global chemical
composition, accurate measurements of the O~K$\alpha$ EWs still
potentially provide a new and relatively direct means of assessing the
veracity of the GS and \citet{Asplund.etal:05} mixtures.

The line EW is also sensitive to the abundances adopted
for the exciting coronal spectrum.  This is due to the large
contribution of the O~VII He-like complex to the source of ionising
photons for coronal temperatures $\log T < 6.5$.  To a lesser extent,
Fe L-shell and Ne, Mg and Si H-like and He-like lines also make a
contribution for temperatures up to $\log T\sim 7.0$.  We have
examined the sensitivity to coronal abundances by comparison of
O~K$\alpha$ line EWs computed for coronal spectra
generated using GS and \citet{Asplund.etal:05} compositions, as
illustrated in Figure~\ref{f:ew}.  The former are higher than the
latter by an amount that decreases from 25-30\%\ for temperatures
$\log T \leq 6.3$ where the O~VII lines dominate, to $\sim 10$\%\ at
$\log T \sim 6.8$--7.0.  Differences at higher temperatures are
largely due to the lower Ne, Mg, Si and Fe abundances in the
\citep{Asplund.etal:05} composition (by 74, 12, 10 and 55\%\,
respectively).

In addition to uncertainties in the solar O content, coronal abundance
variations are also expected as a result of chemical fractionation, in
which the abundances of elements with low first ionisation potentials
are seen to differ from photospheric values by factors of up to $\sim
4$ \citep[e.g.][]{Feldman:92}.  In this context, we emphasise that in
a practical application of the fluorescence technique the photospheric
abundance would be deduced by comparing the model fluorescent
EWs computed for different photospheric abundances and
the {\em simultaneously observed} coronal spectrum, and knowledge of
the actual coronal composition is not required.


\subsection{Observability of O~K$\alpha$ Fluorescence and its
  Abundance Diagnostic Utility} 

As discussed in Paper~II in the context of Ne fluorescence, existing
solar spectra generally have insufficient sensitivity to detect these
relatively weak lines.  While the O~K$\alpha$ line is expected to be
an order of magnitude stronger than that of Ne~K$\alpha$, there are
essentially no high quality solar spectra that reach beyond 23~\AA,
including the most extensive sets of soft X-ray spectra obtained for
the Sun by the satellite-borne Solar Maximum Mission Soft X-ray
Polychromator \citep{Acton.etal:80} and SOLEX
instruments \citep{McKenzie.etal:80} that had coverage up to 22.43 and
23.0~\AA, respectively.  

Existing analyses of higher quality spectra whose ranges encompassed
the O~K$\alpha$ line obtained with rocket-borne instruments
\citep[e.g.][]{McKenzie.etal:78,Acton.etal:85} make no mention of a
feature at the expected wavelength of 23.37~\AA.  The weakest
intensity reported from the spectrum of a flare in the
10-100~\AA\ range obtained by \citet{Acton.etal:85} was
10~ph~arcsec$^2$, which can be compared with the measured O~VII
resonance line intensity of 508~ph~arcsec$^2$.  At a typical solar
coronal temperature of $\log T=6.3$, the flux of the O~VII $\lambda
21.6$ line is a factor $\sim 10^3$ stronger than our prediction for the
O~K$\alpha$ line and the non-detection of the O~K$\alpha$ line is not
a surprise.  While the intensity ratio drops with increasing {\em
  isothermal} temperature to a factor of 10 at $\log T=6.8$, where the
O~VII ion fraction is comparatively small, it would take the presence
of relatively little cooler material within the instrument field of
view to strengthen the observed O~VII $\lambda 21.6$ line
significantly.  Accurate measurement of the O~K$\alpha$ line requires
greater sensitivity than afforded by existing solar spectra.

As shown in \S\ref{s:mcresults}, the O~K$\alpha$ line remains
essentially blend-free and isolated for plasma temperatures
encountered in the non-flaring solar corona.  The line EW
peaks at coronal temperatures of $\log T\sim 6.2$---conditions
typical of those found in the coolest active regions and in the quiet
Sun. In the case of Ne~K$\alpha$, we found fluorescent line fluxes for
quiet Sun conditions of $\sim 0.2$~ph~s$^{-1}$~arcmin$^{-2}$.  The
O~K$\alpha$ line is expected to be an order of magnitude stronger than
this.  Moreover, since the Fe~XXIII and Ca~XV blends remain at levels
significantly below 10\%\ of O~K$\alpha$ for temperatures up to $\log
T\sim 6.5$, the line is also expected to be accessible in much
brighter active regions where fluxes are higher.

As in the case of Ne~K$\alpha$, the large intensity contrast between
O~K$\alpha$ and neighbouring strong lines---in this case the O~VII
He-like complex in the range 21.6-22.1~\AA, and the N~VII $\lambda
24.78$ resonance line---imposes tight instrumental profile
requirements in order to avoid contamination of the O~K$\alpha$ region
from surrounding line wings.  In the absence of suitable simultaneous
spectroscopic observations of the solar regions being studied for
fluorescence, such an instrument would also require fairly broad
wavelength coverage so as to measure accurately the fluorescing X-ray
spectrum from the O~K threshold down to wavelengths at which the
contribution to O~K ionisation is no longer significant.  The short
wavelength cut-off requirement will depend on the coronal temperatures
being studied.

We emphasise that in a practical application of the fluorescence
technique for determining photospheric abundances, knowledge of the 
actual coronal composition is not required since the observed
EW is compared to a measured coronal spectrum that
itself must be used in model predictions.


\section{Conclusions}

We have investigated the observability of the O~K$\alpha$ line formed
in the solar upper photosphere/lower chromosphere by fluorescence from
overlying coronal X-radiation.  While quite faint compared to the
prominent lines of abundant coronal ions, O~K$\alpha$ should be
essentially unblended at plasma temperatures up to $\log T\sim 6.5$.

The line EW peaks at temperatures $\log T\sim 6.2$ with
a value of $\sim 0.1$~\AA, and we estimate an O~K$\alpha$ flux of
$\sim 2$~ph~s$^{-1}$~arcmin$^{-2}$ for the quiet Sun at heliocentric
angles close to $0^\circ$---an order of magnitude larger than
predicted in Paper~II for Ne~K$\alpha$ fluorescence.  Both lines
should be quite observable with instruments adequately tailored toward
high resolution $\lambda/\Delta\lambda \ga 1000$ and sufficiently
well-behaved instrumental profile wings.

The O~K$\alpha$ line provides a potentially accurate absolute O abundance
diagnostic, provided abundances of other prominent light elements are
also constrained.  It should also be possible to distinguish between
the currently competing photospheric compositions of GS and
\citet{Asplund.etal:05}, provided line EWs can be
determined to an accuracy significantly better than 20\%.  We conclude
that X-ray fluorescence of O (and Ne) represents a viable means for
resolving the current solar abundance-helioseismology conundrum.

\acknowledgments

We thank the NASA AISRP for providing financial assistance for the
development of the PINTofALE package.  JJD was funded by NASA contract
NAS8-39073 to the {\em Chandra X-ray Center} during the course of this
research and thanks the Director, H.~Tananbaum, for continuing support
and encouragement.  BE was supported by {\it Chandra} Grants GO6-7008X
and GO6-7098X.  Finally, we thank the anonymous referee for comments that
helped us clarify the manuscript.



\newpage

\begin{table}
\label{t:5sigma}
\caption{Lines in the CHIANTI 5 database lying within $\pm 0.07$~\AA\ 
($5\sigma$ + a rest wavelength uncertainty 0.02~\AA) of the
  position of the 23.37~\AA\ O~K$\alpha$ transition for a Gaussian
  resolving power of $\lambda/\Delta\lambda=1000$.  Only lines with
  predicted intensities $\ge 1\times 10^{-4}$ that of the brightest
  line are listed.} 
\begin{tabular}{lcccc}
\hline
Ion & $\lambda$ (\AA) & Rel. Int. & $\log T_{max}$ & Transition \\
\hline

  Fe {\sc XXIII} &  23.303 & 0.54 & 7.2 & $2s3p \; ^3P_{2}$ --   $2s5d \;
  ^3D_{3}$   \\

     Ca {\sc XV} &  23.305 & 0.07 & 6.6 & $2s^22p^2 \; ^1D_{2}$  -- $2s^22p^3d \; ^3D_{1}$  \\

    Ca {\sc XVI} &  23.320 & 0.03 & 6.7 & $2s2p^2 \; ^2P_{3/2}$ --
    $2s2p(^3P)3d \; ^4D_{5/2}$   \\

  Fe {\sc XXIII} &  23.321 & 0.10 & 7.2 & $2s3p \; ^3P_{2}$  -- $2s5d \;
  ^3D_{2}$   \\

    Ca {\sc XVI} &  23.341 & $1\times 10^{-3}$&  6.7 & $2s2p^2 \;
    ^2P_{3/2}$  -- $2s2p(^3P)3d \; ^4D_{3/2}$   \\

    Ar {\sc XVI} &  23.343 & 0.01 & 6.7 & $1s^22s \; ^2S_{1/2}$  -- $1s^23d
    \; ^2D_{5/2}$ \\

    Ar {\sc XVI} &  23.362 & 0.01 & 6.7 & $1s^22s \; ^2S_{1/2}$  -- $1s^23d
    \; ^2D_{3/2}$   \\

  Fe {\sc XXIII} &  23.363 & 1.00 & 7.2 & $2s3p \; ^3P_{1}$  -- $2s5s \;
  ^1S_{0}$   \\

  Fe {\sc XXIII} &  23.381 & 0.04 & 7.2 & $2s3p \; ^1P_{1}$  -- $2s5d \;
  ^3D_{1}$   \\

  Fe {\sc XXIII} &  23.388 & $2\times 10^{-4}$ & 7.2 & $2s3p \;
  ^1P_{1}$  --
  $2s5p \; ^3P_{1}$   \\

  Fe {\sc XXIII} &  23.388 & $3\times 10^{-4}$ & 7.2 & $2s3p \;
  ^1P_{1}$  --
  $2s5p \; ^1P_{1}$   \\

     Ca {\sc XV} &  23.388 & 0.88 & 6.6 & $2s^22p^2 \; ^1S_{0}$  --
     $2s^22p^3d \; ^1P_{1}$  \\

  Fe {\sc XXIII} &  23.406 &  $8\times 10^{-3}$ & 7.2 & $2s3p \; 
  ^3P_{2}$ -- $2s5d \; ^3D_{1}$ \\

  Fe {\sc XXIII} &  23.412 & $3\times 10^{-4}$ &  7.2 & $2s3p \; 
  ^3P_{2}$ -- $2s5p \; ^3P_{1}$ \\

     Ca {\sc XV} &  23.415 &    0.45 & 6.6  & $2s^22p^2 \; ^1D_{2}$ -- 
     $2s^22p^3d \; ^3F_{3}$ \\

 Ni {\sc XXVIII} &  23.439 & $5\times 10^{-4}$ & 8.0 & $3p \; 
 ^2P_{1/2}$ -- $4d \; ^2D_{3/2}$ \\

\hline
\end{tabular}
\end{table}

\newpage

\end{document}